\documentclass[emulateapj]{emulateapj}
\usepackage{rotating}
  \received{}
  \revised{}
  \accepted{}

\shortauthors{Napolitano et al.} \shorttitle{$f(R)$-gravity in
elliptical galaxies}

\def\Re{\mbox{$R_{\rm eff}$}}

\def\kms{\mbox{km s$^{-1}$}}

\def\lsim{\mathrel{\rlap{\lower3.5pt\hbox{\hskip0.5pt$\sim$}}
    \raise0.5pt\hbox{$<$}}}                
\def\gsim{~\rlap{$>$}{\lower 1.0ex\hbox{$\sim$}}}

\begin{document}

\title{Testing Yukawa-like potentials from $f(R)$-gravity in elliptical galaxies}

\author{N.R.~Napolitano\altaffilmark{1}, S. Capozziello\altaffilmark{2,3}, A.J.~Romanowsky\altaffilmark{4},
M. Capaccioli\altaffilmark{2}, C.~Tortora\altaffilmark{5}}


\altaffiltext{1}{INAF -- Osservatorio Astronomico di Capodimonte,
Salita Moiariello, 16, 80131 - Napoli, Italy}

\altaffiltext{2}{Dipartimento di Scienze Fisiche, Universit\`a di Napoli
``Federico II'', Napoli, Italy}

\altaffiltext{3}{Istituto Nazionale di Fisica Nucleare, Sez. di Napoli, Italy}

\altaffiltext{4}{UCO/Lick Observatory, University of California,
Santa Cruz, CA 95064, USA}

\altaffiltext{5}{Universit$\rm \ddot{a}$t Z$\rm \ddot{u}$rich,
Institut f$\rm \ddot{u}$r Theoretische Physik, Winterthurerstrasse
190, CH-8057, Z$\rm \ddot{u}$rich, Switzerland}\email{\texttt
napolita@na.astro.it}

\begin{abstract}
We present the first analysis of extended stellar kinematics of
elliptical galaxies where a Yukawa--like correction to the Newtonian 
gravitational potential derived from $f(R)$--gravity is considered as
an alternative to dark matter. In this framework, we model long-slit 
data and planetary nebulae data out to 7 \Re\ of 
three galaxies with either decreasing or flat dispersion profiles. 
We use the corrected Newtonian potential 
in a dispersion--kurtosis Jeans analysis to account for the
mass--anisotropy degeneracy. We find that these modified potentials
are able to fit nicely all three elliptical galaxies and
the anisotropy distribution is consistent with that estimated if a 
dark halo is considered.
The parameter which measures the ``strength'' of the Yukawa--like 
correction is, on average, smaller than the one found previously in 
spiral galaxies and correlates both with the scale length of the
Yukawa--like term and the orbital anisotropy. 
\end{abstract}

\keywords{galaxies : kinematics and dynamics  --  galaxies : general -- 
galaxies : elliptical and lenticular -- cosmology: theory}

\section{Introduction} \label{sec:intro}
The ``concordance'' $\Lambda$CDM cosmological model, which
includes some unseen Cold Dark Matter (DM) and a cosmological constant
($\Lambda$) acting as a repulsive form of Dark Energy (DE), has
been remarkably successful in explaining the  formation
and evolution of cosmological structures at different scales (e.g., \citealt{2006Natur.440.1137S}).

However, at cosmological scales, 
the cosmological constant as ``vacuum state'' of the gravitational
field is about 120 orders of magnitude smaller 
than the value predicted by any quantum gravity
theory (\citealt{weinberg}) and comparable 
to the matter density (coincidence problem), even if
they evolved decoupled in the history of the universe. 

In addition, looking at the galaxy scales there are a few critical issues
yet to be solved, which are giving hard time to the whole
$\Lambda$CDM framework. 

Since the discovery of the flat rotation curves 
of spiral systems, galaxies have been the most critical laboratory 
to investigate the gravitational effects of the DM halos, 
to be compared against the
expectation of the cosmological simulations
(\citealt{nfw97}, NFW hereafter, \citealt{1995ApJ...447L..25B}, \citealt{2010MNRAS.402...21N}). Here, the
$\Lambda$CDM model is not able to fully explain the shallow
central density profile of spiral and dwarf galaxies
\citep{2007ApJ...663..948G,2007MNRAS.378...41S,2008ApJ...676..920K}.
Early-type galaxies (ETGs hereafter) have been proven only recently 
to be consistent with $\Lambda$CDM predictions (and WMAP5
cosmological parameters, e.g. \citealt{2009ApJS..180..330K}), from their 
centers (\citealt{2010MNRAS.405.2351N})
to their peripheries (\citealt{2011MNRAS.411.2035N}, N+11
hereafter), although there are also diverging results showing that ETGs
in some cases have too high 
(\citealt{2007ApJ...664..123B}) or too low (e.g.
\citealt{2008JCAP...08..006M}) concentrations.


This very uncertain context has been a fertile soil for alternative
approaches to the so-called ``missing mass''. The basic approach
is that the Newtonian Theory of Gravity, which has been tested
only in the Solar System, might be inaccurate on larger (galaxies
and galaxy clusters) scales. The most popular theory investigated
so far, the Modified Newtonian dynamics (MOND) proposed by
\citet{1983ApJ...270..365M}, is based on phenomenological
modifications of Newton dynamics in order to explain the flat
rotation curves of spiral galaxies, and passed a number of 
observational tests (\citealt{2009Sci...326..812F}),
included ETG kinematics (\citealt{2003ApJ...599L..25M}; 
\citealt{2007A&A...476L...1T}; \citealt{2010A&A...523A..32K}; 
\citealt{2010arXiv1011.5741C}; \citealt{2011A&A...531A.100R}). 
Only lately it has been
derived in a cosmological context \citep{2004PhRvD..70h3509B}.

A new approach, motivated from cosmology and quantum field 
theories on a curved space-time, has been proposed to study the
gravitational interaction: the Extended Theories of Gravity  
(\citealt{capozz+02}; \citealt{capozzfara}). 
In particular, the so called $f(R)-$gravity seem to have passed
different observational tests like spiral galaxies' rotation
curves, X-ray emission of galaxy clusters and 
cosmic acceleration  (see e.g. \citealt{capozz+07}, C+07 hereafter, 
\citealt{capozz+09}, C+09 hereafter, \citealt{capozz+08}). 
This approach is based on a straightforward generalization of Einstein theory where 
the gravitational action (the Hilbert-Einstein action)  is assumed to 
be linear in the Ricci curvature scalar $R$. In the case of $f(R)$-gravity, 
one assumes a generic function $f$ of the Ricci scalar $R$ 
(in particular analytic functions) and asks for a theory of gravity having 
suitable behaviors at small and large scale lengths.  

As shown in \citet{stabile2}, analytic $f(R)$-models give rise, in general, 
to Yukawa--like corrections to the Newtonian potentials in the weak field limit 
approximation (see also \citealt{arXiv:1104.2851}). The correction introduces a new gravitational scale, besides 
the standard Schwarzschild one, depending on the dynamical structure of the 
self-gravitating system.

Here we want to test these 
Yukawa-like gravitational  potentials against a sample of elliptical galaxies. 
This approach has been   proposed earlier, in a phenomenological scheme for
anti--gravity,  to model flat rotation curves of spiral galaxies 
\citep{1984A&A...136L..21S}, and recently, in $f(R)$ theories to model 
disk galaxies combined with NFW haloes (see \citealt{cardone11}).
The test we are proposing at galaxy scales is crucial: reproducing kinematics 
and then dynamics of these very different classes of astrophysical systems in 
the realm of the same paradigm is needed to test 
these new gravitational theory as an alternative to DM which has 
not been definitely found out at fundamental level.

The layout of the paper is the following.  In \S 2, we sketch the main ingredients of $f(R)$-gravity deriving, in the weak field limit, the Yukawa-like corrected gravitational potential. \S 3 is devoted to the high-order Jeans analysis suitable for ellipticals.  The dispersion-kurtosis fitting and the data sample are presented in \S 4. Discussion and conclusions are in \S 5.


\section{Post- Newtonian  potentials from $f(R)$-gravity}\label{sec:sec2}

We are interested in testing a class of modified potentials which
naturally arise in post-Newtonian approximation of $f(R)-$gravity
for which no particular choice of the Lagrangian has been 
provided.

The starting point is a general gravity action of the form

\begin{eqnarray}\label{actfR}
\mathcal{A}\, = \,\int
d^4x\sqrt{-g}\biggl[f(R)+\mathcal{X}\mathcal{L}_m\biggr]\,,
\end{eqnarray}
where $f(R)$ is an  analytic function of Ricci scalar, $g$ is the
determinant of the metric $g_{\mu\nu}$, ${\displaystyle \mathcal{X}=\frac{16\pi
G}{c^4}}$ is the gravitational coupling constant, and $\mathcal{L}_m$ describes
the standard fluid-matter Lagrangian. Such
an action is the straightforward generalization of the
Hilbert-Einstein action   obtained as soon as  $f(R)=R$. 

In \citet[and reference therein]{stabile2} it has been shown that if one solves the field equations 
in the weak field limit under the general assumption of 
an analytic Taylor expandable $f(R)$ functions of the form

\begin{eqnarray}\label{sertay}
f(R)\simeq
f_0+f_1R+f_2R^2+f_3R^3+...\,
\end{eqnarray}
 
the following gravitational potential arises

\begin{eqnarray}\label{gravpot}
\Phi\,=\,-\left(\frac{GM}{f_1r}+\frac{L\delta_1(t)~e^{-\frac{r}{L}}}{6~
r}\right) ,
\end{eqnarray}

where $L\doteq\displaystyle-\frac{6f_2}{f_1}$, $f_1$ and $f_2$
are the expansion coefficients obtained by Taylor expansion.  
We note that the $L$ parameter is related to the effective mass $m = (
-3/L^2)^{-1/2}=(2f_2/f_1)^{1/2}$ and can be interpreted also as an effective length.

From Eq. \ref{gravpot}, the standard Newton potential is
recovered only in the particular case $f(R)=R$. 
Furthermore, the parameters $f_{1}$ and $f_{2}$ and the function
$\delta_1$ represent the deviations with respect the standard
Newton potential. On the Solar system scale, it has been shown 
that Yukawa--like deviation from the pure 
Newtonian potential are not in contradiction with classical tests of General 
Relativity (see e.g. \citealt{2008PhRvD..77j7501C}), thanks to the 
so-called Chameleon mechanism (\citealt{2004PhRvD..69d4026K}). In 
particular, $f_{1}$ and $f_{2}$
parameters are expected to allow the regular Newtonian
potential, while at larger scales they can assume non-trivial values (e.g.
$f_1\neq 1,\,\delta_1(t)\neq 0,\,\xi\neq 1$, see \citealt{2007PhRvD..76j4019C}, \citealt{stabile2}).

Eq. (\ref{gravpot}) can be recast as 
\begin{equation}
\label{gravpot1} \Phi(r) = -\frac{G M}{
(1+\delta) r}\left(1+\delta e^{-\frac{r}{L}}\right)\,,
\end{equation}
where the first term is the Newtonian--like part of the potential
associated to baryonic point--like mass $M/(1+\delta)$ (no DM) and the second
term is a modification of the gravity including a ``scale
length'', $L$ associated to the above coefficient of the Taylor expansion.
If $\delta=0$ the Newtonian potential is recovered. 
Comparing Eqs. \ref{gravpot} and \ref{gravpot1}, we obtain that 
$1+\delta=f_1$, and $\delta$ is related to $\delta_1(t)$ 
through
\begin{equation}\label{eq:delta1}
\delta_1=-\frac{6GM}{L^2}\frac{\delta}{1+\delta}
\end{equation}
where $6GM/L^2$ and $\delta_1$ can be assumed quasi-constant.  
From Eq. \ref{eq:delta1}, it turns out that 
$L\propto \sqrt{- \delta/(1+\delta)}$.
Due to the arbitrarity of $\delta_1(t)$, the actual value
of the $\delta$ parameter can assume any values, however, in order
to have a Yukawa potential with a non imaginary exponent (i.e.
$L$ must be real) it
is required that $\xi<0$ or $-1<\delta<0$. 
As comparison, \citet{1984A&A...136L..21S}
adopted the same potential as in Eq. \ref{gravpot1} under the 
assumption of anti--gravity generated by massive particles (of mass $m_0$) 
carrying the additional gravitational force. 
In this case a typical scale length would naturally
arise ($L=h / m_0 c$ being a Compton length) and  a 
$-1 < \delta < 0$ would provide
a repulsive term to the Newtonian--like term producing flat 
rotation curves at $r\gg L$ as observed in spiral galaxies.
In particular, for a small sample of spiral systems \citet{1984A&A...136L..21S} 
found $-0.95\lsim \delta\lsim -0.92$.


\begin{figure}
\centering 
\vspace{-0.45cm}
{\hspace{-0.55cm}}\includegraphics[width=0.51\textwidth,clip]{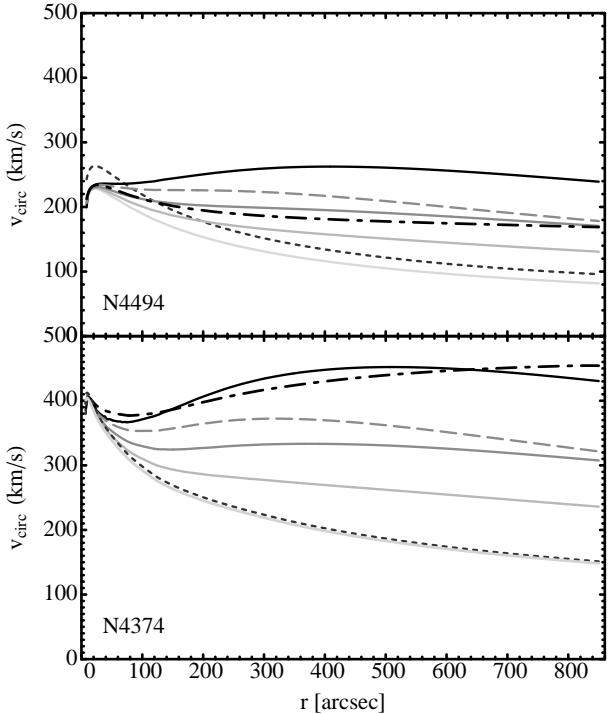}
\vspace{-0.7cm}
\caption{Circular velocity produced by the modified potential in 
Eq. \ref{gravpot1} for the two galaxies N4494 (top) and N4374 (bottom). 
In both cases the $M/L_*$ has been fixed to some fiducial value (as
expected from stellar population models and Kroupa 2001 IMF): $M/L_*=4.3 
\Upsilon_{\odot, B}$
for NGC 4494 and $M/L_*=5.5\Upsilon_{\odot, V}$ for NGC 4374.
The potential parameters adopted are: $L= 250''$ and $\delta$=0, -0.65, -0.8, -0.9 (lighter to darker solid lines) and $L=180''$ and $\delta$=-0.8 (dashed lines). 
The dotted line is a case with positive coefficient of the Yukawa--like
term and $L=5000''$ which illustrates that positive $\delta$ cannot produce
flat circular velocity curves. Finally some reference Navarro--Frenk--White (NFW)  models are shown as 
dot-dashed lines.
}\label{fig:fig0}
\end{figure}

Here we want to test the modified potential as in Eq. \ref{gravpot1} in 
elliptical galaxies and check whether it is able to provide a
reasonable match to their kinematics and how the model parameters
compare with the results obtained from spiral systems.
We construct equilibrium models based on the solution of the
radial Jeans equation (see \S\ref{sec:jeans}) to interpret the
kinematics of planetary nebulae (PNe, see \citealt{nap+02,nap+05};
\citealt{2003Sci...301.1696R}; \citealt{2009MNRAS.394.1249C}) 
which are the {\it only stellar--like tracers for galaxy dynamics available in
ETGs out to $\sim5-10$ effective radii} (\Re). 

We will use the inner long slit data and the extended PN kinematics for
three galaxies which have published dynamical analyses within 
DM halo framework: NGC 3379 (\citealt{2007ApJ...664..257D}; 
\citealt{2009MNRAS.395...76D},
DL+09 hereafter), NGC 4494 (\citealt{nap+09}, N+09), NGC 4374
(N+11). 
The decreasing velocity dispersion profiles of the first
two galaxies have been modeled with an intermediate mass halo,
$\log M_{\rm vir}\sim 12-12.2 M_\odot$, with concentration $c_{\rm
vir}=6-8$ and a fair amount of radial anisotropy in the outer
regions. For NGC 4374, having a rather flat dispersion
profile, a more massive (adiabatically contracted) halo with $\log
M_{\rm vir}\sim 13.4 M_\odot$ and $c_{\rm vir}\sim7$ was required
with a negligible amount of anisotropy in the outer regions. 
These models turned out to be in fair agreement with the
expectation of WMAP5 $c_{\rm vir}-M_{\rm vir}$ relation
(N+11) and with a
\citet{2001MNRAS.322..231K} IMF, making this sample particularly
suitable for a comparison with an alternative theory of gravity
with no-DM as we want to propose here.

Before we go on with detailed stellar dynamics, we show in Fig. \ref{fig:fig0}
the circular velocity of the
modified potential as a function of the potential parameters $L$ and
$\delta$ for NGC 4494 and NGC 4374.
As for the spiral galaxies, negative values of the $\delta$ parameter make the
circular velocity more and more flat also reproducing the typical dip 
(e.g. NGC 4374) of the circular velocity found for the DM models 
(dot--dashed curves) of the most massive systems. On the contrary, 
positive $\delta$ values cannot produce flat circular velocity curves (see Fig. 
\ref{fig:fig0}).  

\begin{figure*}
\centering 
\vspace{-0.35cm}
\hspace{-0.5cm}
\includegraphics[width=1.\textwidth,clip]{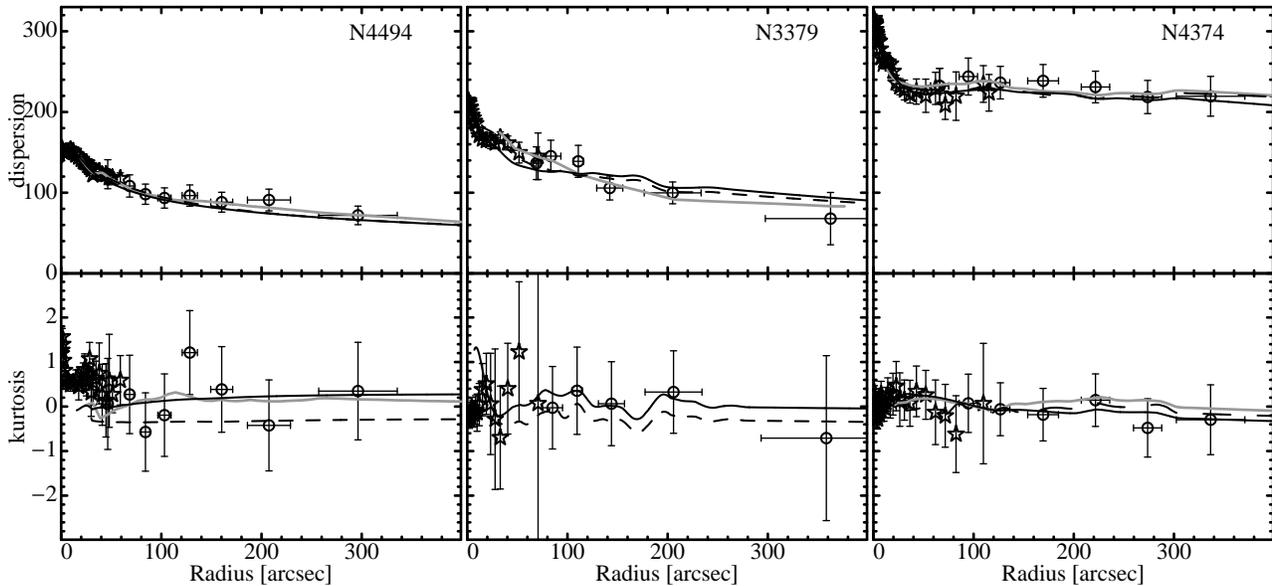}
\vspace{-0.6cm}
\caption{Dispersion in \kms\ (top) and kurtosis (bottom) fit of the galaxy sample
for the different $f(R)$ parameter sets: the anisotropic solution
(solid lines) is compared with the isotropic case (dashed line -- for NGC 4374 and 
NGC 4494 this is almost indistinguishable from the anisotropic case). 
From the left, NGC 4494, NGC 3379 and NGC 4374 are shown with
DM models as gray lines from N+09, DL+09 
(no kurtosis is provided), and N+11
respectively.}\label{fig:fig1}
\end{figure*}


\subsection{A consistency check with galaxy scaling relations}
To conclude the inspection of the modified potential as in Eq. \ref{gravpot1}
here we want to show that, beside flat rotation curves, this also naturally 
accounts for fundamental scaling relations of galaxies: the Tully-Fisher (TF) 
relation for spirals and Faber-Jackson (FJ) relation for ETGs.

Both relations connect the total mass $M$ of galaxies with some characteristic
velocity defining the kinetic energy of the systems (i.e. the 
maximum rotation velocity, $v_{\rm max}$, for spirals and the central velocity 
dispersion, $\sigma_0$, for ETGs). In either cases the kinematical 
quantities involved are proportional to the circular velocity of the systems 
through some ``structure'' constant, thus the arguments below apply
to galaxies in general.
 
Although the point--like version of the potential
implies that the circular velocity $v_c$ scales with mass as $M \sim v_{\rm c,max}^2$
(as pointed out by \citealt{1984A&A...136L..21S}), if one derives the circular
velocity for an extended galaxy this can be generalized as
\begin{equation} 
v_c^2(r) = (GM_{\rm tot}/r_*) \times f(r/r_*; \delta, L/r_*)
\end{equation} 
where $r_*$ is a characteristic radius (e.g. the disk length for spirals or
the effective radius encircling half of the galaxy light for ETGs), 
$f(r/r_*;~\delta,~L/r_*)$ is a generic function which 
includes the radial dependence of the enclosed mass and the above Yukawa-like 
term. This function is defined such as, for $\delta=0$, it gives 
$v_c^2(r) = GM(r)/r$ as the usual Newtonian expression. 
It is easy to show that if galaxies are homologous the maximum of $v_c$ 
is reached at the same $r/r_*$, for a given $\delta$ and $L/r_*$ and 
this maximum can be written as 
\begin{equation}\label{eq:vmax}
v_{\rm c,max}^2 = {\rm K} M_{\rm tot}/r_*
\end{equation}
where the constant $K$ depends on the set of parameters \{$\delta, ~L, ~r_*$\} 
adopted. In Eq. \ref{eq:vmax}, though, $M_{\rm tot}$ and $r_*$ are linked by
the size--mass relation which is generally written as
$r_* \propto M_{\rm tot}^\alpha$, 
from which Eq. \ref{eq:vmax} can be written as
\begin{equation}\label{eq:vmax2}
v_{\rm c,max}^2 \propto M_{\rm tot}^{1-\alpha}.
\end{equation}
The size-mass relation of spiral galaxies can be found in  
\citet[see also \citealt{2009ApJ...691..770T}]{1996MNRAS.281...27P} to be
$r_*\propto M_{\rm tot}^{0.4}$, while it is $r_*\propto M_{\rm tot}^{0.6}$
for ETGs (e.g. \citealt{2003MNRAS.343..978S}, \citealt{nap+05}). 
This would give a TF slope of $3.33$ and FJ slope of $5$ which are both in 
the range of the observed relations (see e.g. \citealt{2005ApJ...632..859M}
and \citealt{2010A&A...516A..96N} respectively) with the remaining discrepancy 
being mainly due to the conversion
factor to the observed quantities and non homologies.
 
We finally remark that the TF relation has been found not to be
conflicting with $f(R)$ potentials in \citet{2006JCAP...08..001C}, although 
the potentials from $f(R)\propto R^n$ adopted there are just a series expansion of 
the  Yukawa--like potential coming out from a more general polynomial f(R) as in 
Eq. \ref{gravpot}.

\section{High-order Jeans analysis}\label{sec:jeans}
From the model point of view, the problem of fitting a modified
potential as in Eq. \ref{gravpot1} (which is formally
self-consistent since the source of the potential is the only mass of
the dynamical tracers, i.e. stars) implies the same kind of
degeneracies between the anisotropy parameter, $\beta=1-\sigma_\theta^2/\sigma_r^2$ 
(where $\sigma_\theta$ and $\sigma_r$ are the azimuthal and
radial dispersion components in spherical coordinates), and the 
non--Newtonian part of the potential (characterized by two parameters
like typical dark haloes) in a similar way of the classical
mass-anisotropy degeneracy. We have shown (N+09, N+11) that these
degeneracies can be alleviated via higher-order Jeans equations including 
in the dynamical models both the
dispersion\footnote{For the
slow--rotating models we use the {\em} velocity, $v_{\rm
rms}=\sqrt{v^2+\sigma^2}$ as a measure of the velocity
dispersion.} ($\sigma_{\rm p}$) and the kurtosis ($\kappa$) profiles of the tracers.

In the following, we will use the assumption of spherical symmetry since
galaxies in the sample 
are all E0--E1 for which, if one exclude the singular
chance that they are all flattened systems seen face--on (see
discussion in Sect. 8.1. of \citealt{2007ApJ...664..257D}), the
spherical approximation is good at 10\% (\citealt{2000A&AS..144...53K})\footnote{
The effect of non--spherical models is outside the scope of this paper, 
but details for NGC 3379 and NGC 4494 can be found in DL+09 and N+09.}.
Under spherical assumption, 
no-rotation and $\beta=\rm const$ (corresponding to the family 
of distribution functions $f(E,L)=f_0 L^{-2\beta}$, see 
\citealt{2002MNRAS.333..697L} and references therein\footnote{ Here, there is
the caveat that the solution 
of Jeans Equations does not ensure that the final 
distribution function is non negative and thus fully physical (see e.g., 
\citealt{2006AJ....131..782A}).}, the 2-nd and 4-th moment radial equations 
can be compactly written as:
\begin{equation}\label{eq:jeans2-4}
s(r)= r^{-2\beta}\int_r^\infty x^{2\beta} H(x) dx
\end{equation}

where $s(r)=\{\rho\sigma_r^2; \rho \overline{v_r^4}\}$, $\beta$ is
the anisotropy parameter, and

$$H(r)=\left\{ \rho\frac{d\Phi}{dr};
3\rho\frac{d\Phi}{dr}\overline{v_r^2} \right\} $$

respectively for the dispersion and kurtosis equations, being the
latter $\kappa(r)={\overline{v_r^4}}/\sigma_r^4$. In the same equations, 
$\Phi(r)$ is the spherical extended 
source version of the point--like potential as in Eq.
\ref{gravpot1}\footnote{This is obtained assuming the onion
shell approximation:
$\Phi(r)=\int_0^r\int_0^{2\pi}\int_0^\pi\phi(r) ~r^2\sin\theta~
 d\theta d\varphi dr$, see also Eq. 18 of C+09.}
 and $\rho(r)$ is the 3D density of the tracer obtained by multiplying the 
deprojection of
the stellar surface brightness profile, $j_\star(r)$, by some
constant stellar mass-to-light ratio, $M/L_\star$.  

This $M/L_\star=const$
might be a strong assumption to check further in a separate paper
as it neglects the presence of stellar population gradients (see
e.g. Tortora et al. 2010). However, colour (and $M/L$) gradients are generally
stronger within \Re\ (see e.g. \citealt{tortora+10}) and might
mainly drive the best fit in the central regions, while they are possibly 
shallower outside (e.g., \citealt{2003AJ....126..596T}) where the $f(R)$
parameters should be better constrained.
In the following, $j_\star(r)$ is derived by photometry presented 
in previous dynamical studies (i.e. DL+09, N+09,
N+11 for NGC 3379, NGC 4494 and NGC 4374 respectively).
 
  
Eqs. \ref{eq:jeans2-4}  are the ones
interested by the potential modification and include four free
parameters to be best-fitted: the $f(R)$ parameters \{$\delta,
L$\}, the ``dynamically inferred'' stellar mass-to-light ratio
$M/L_\star$ and the constant anisotropy $\beta$ (see also
\S\ref{sec:results}). The solutions of Eqs. \ref{eq:jeans2-4} on a
regular grid in the parameter space are then projected to match
the observed line-of-sight kinematical profile via ordinary Abel
integrals (see N+09 for details).

As mentioned earlier, Eqs. \ref{eq:jeans2-4} are written under the
assumption of a constant $\beta$ with radius, which provides a
average global anisotropy distribution over all the galaxy. As
seen in previous analyses (e.g. N+09, DL+09 and N+11), it is
likely that this might not be a fair assumption, as $\beta$ turns out
to be constant somewhere in the outer regions, but strongly varying
in the central radii.
In this preliminary test we will skip this implementation of the
models since we expect this to possibly improve the fit to the
data in the central part only, where we do not expect the overall
dynamics of being strongly ruled by the $f(R)$ potential, whose
parameters are the main focus of this work. Furthermore, we
have shown previously (see e.g. N+09 and N+11) that the assumption
of the constant or radial varying anisotropy did not strongly
affect the determination of the other important parameter, the
dynamically based stellar $M/L$. In the following we will take the
$\beta=$const as fair estimate of the average galaxy anisotropy.

\section{Dispersion-kurtosis fitting}\label{sec:results}
In Fig. \ref{fig:fig1} we show the dispersion and kurtosis
profiles of the three galaxies with the $f(R)$ models
superimposed (solid lines). The fitting procedure is based on the simultaneous
$\chi^2$ minimization of the dispersion and kurtosis profiles over
a regular grid in the parameter space. The best--fit parameters
are summarised in Table \ref{tab:tab1} together with some info of
the galaxy sample.

Overall the agreement of the model curves with data is remarkably
good and it is comparable with models obtained with DM modeling
(gray lines in Fig. \ref{fig:fig1}).

In all cases, the $f(R)$ models allow to accommodate a constant 
orbital anisotropy $\beta$ which is very
close to the estimates from the DM models (see e.g. Table
\ref{tab:tab1}\footnote{For NGC 4374 only to be nicely
fitted at all radii we needed to include some radial anisotropy
in the very central regions, following the $\beta(r)$ distribution 
adopted in N+11 (see Eq. 5, whith best fit $r_a=22.5$).}). 
This is mainly guaranteed by the fit to the $\kappa(r)$ which does
not respond much differently to the modified potential with
respect the DM models. Thus, an important result of the analysis
is that the orbital anisotropy is fairly stable to the change of
the galaxy potential. 
In particular, the use of the kurtosis profiles has allowed us 
to solve the degeneracy of the models and favor the anisotropic solutions 
for NGC~3379 and NGC~4494 (NGC~4374 being almost isotropic everywhere). 
Although the isotropy solutions provide also a good fit 
for the dispersion  profile only (see e.g. the dashed lines in Fig. 
\ref{fig:fig1}), they not correctly match the observed $\kappa$.
This produces a significantly worse total $\chi^2$/dof (NGC 3379: 45/26; 
NGC4374: 35/40; NGC 4494: 27/44) with respect to the best--fit in the Table \ref{tab:tab1}, 
although still close to $\chi^2$/dof$~\sim 1$ mainly because of the 
large error bars.

\begin{table*}\label{tab:tab1}
\caption{Model parameters for the $f(R)$ potential.}
\hspace{-0.9cm}\vspace{-0.9cm} \label{tab:tab1}
\begin{center}
\noindent{\smallskip}
\begin{tabular}{llllllll}
\hline \hline Galaxy& Mag (band) & \Re
&$M/L_\star$ & $L$ & $\delta$ & $\beta$ & $\chi^2$/dof\\
\hline
NGC3379  & -19.8(B) & 2.2 &  6 (7) & 6 & -0.75 & 0.5($<$0.8) & 14/25\\
NGC4374  & -21.3(V) & 3.4 &  6 (6) & 24 & -0.88 & 0.01(0.01)& 14/39\\
NGC4494  & -20.5(B) & 6.1 &  3 (4)  & 20 & -0.79 & 0.5(0.5)& 18/43\\
\hline
\end{tabular}
\noindent{\smallskip}\\
\begin{minipage}{10.5cm}
NOTES -- Galaxy ID, total magnitude, effective radius and model
parameters for the unified solution. DM--based estimates for
$M/L_\star$ and $\beta$ (NGC~3379: DL+09; NGC~4374: N+11; NGC~4494: N+09) 
are shown  in parentheses for
comparison. $M/L_\star$ are in solar
units, \Re\ and $L$ in kpc. Typical errors on $M/L_\star$ are of the order 
of 0.2$M/L_\odot$ and on $\beta$ of 0.2 (see also Fig. \ref{fig:fig2}). 
The small $\chi^2$ values are mainly due to the large data error bars.
\end{minipage}
\noindent{\smallskip}\\
\end{center}
\end{table*}

Finally, 
the best fit $M/L_\star$ in Table \ref{tab:tab1} are very similar to the 
values found for DM models 
(reported between brackets) in all cases, generally consistent with
a Kroupa (2001) IMF.

\begin{figure}
\centering 
\vspace{-0.2cm}
\includegraphics[width=0.51\textwidth,clip]{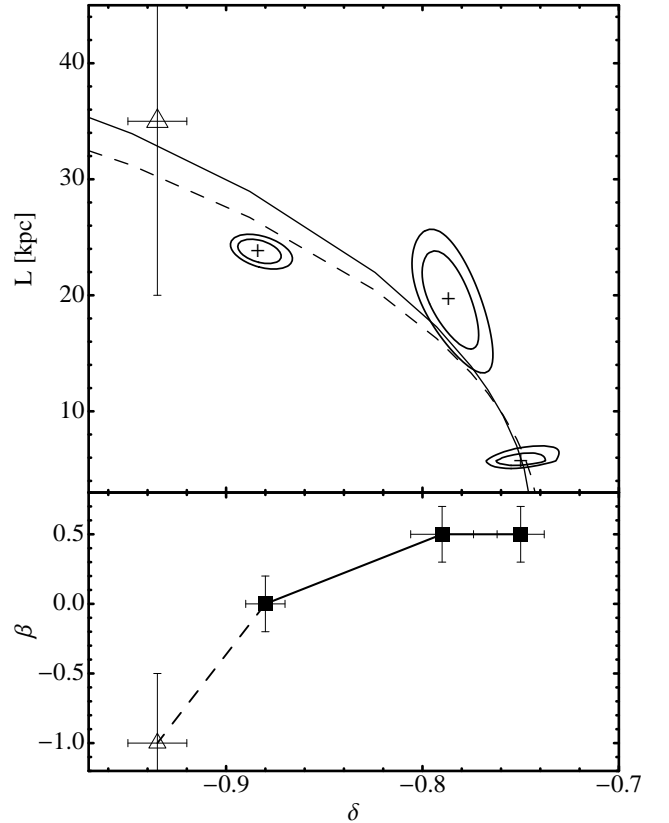}
\vspace{-0.8cm}
\caption{Top: 1- and 2-$\sigma$ confidence levels
in the $\delta-L$ space marginalized over $M/L_\star$ and
$\beta$ (see also Table \ref{tab:tab1}). Spiral galaxy 
results from \citet{1984A&A...136L..21S} are shown as empty triangle with
error bars. Solid (dashed)
curve shows the tentative best-fit to the data 
including (excluding) the spiral galaxies and assuming
a $L\propto \sqrt{\delta/(1+\delta)}$ correlation as expected from
Eq. \ref{eq:delta1}. Bottom: the anisotropy and the $\delta$
parameters turn out to be correlated for the elliptical sample (full
squares). This correlation seems to include also the spiral
sample cumulatively shown as the empty triangle (here we have assumed
$\beta=-1.0\pm0.5$ as a fiducial value for spiral galaxies to draw
a semi-quantitative trend across galaxy types).}\label{fig:fig2}
\end{figure}

Looking at the $f(R)$ parameters, in Fig. \ref{fig:fig2} we
show the marginalized confidence contours of the main two potential
parameters for the three galaxies. As also reported in Table \ref{tab:tab1},
the $\delta$ parameter has a mean value $\delta=-0.81\pm0.07$ which is 
inconsistent with the one previously found
for spiral galaxies (e.g. \citealt{1984A&A...136L..21S},
also shown in Fig. \ref{fig:fig2}). 
On the contrary, $\delta$ seems nicely correlated with the other potential 
parameter, $L$, as expected from Eq. \ref{eq:delta1}. 
In the same figure the correlation is supported by the tentative fit into 
the $\delta-L$ plane (whether or not the spiral galaxy sample is included in the fit), 
although the sample is too small to drive any firm conclusion.  

Interestingly, there seems to be a 
possible increasing trend of $\delta$ with the orbital 
anisotropy: this is also shown in Fig. \ref{fig:fig2} 
where we have added the fiducial value
obtained for the spiral sample (having assumed a reference 
$\beta=-1$ for fiducial tangential anisotropy for late-type systems, 
see e.g. \citealt{2005MNRAS.364..433B}).
This evidence
leaves room for an interpretation of $\delta$ 
and the physics of the galaxy collapse (e.g. the spherical infall
model, \citealt{1972ApJ...176....1G}; \citealt{1977ApJ...218..592G}).

In fact, as discussed in \S\ref{sec:sec2},  
$\delta$ is linked to $\delta_1$, 
which is an arbitrary function that 
comes out because the field equations in the post-Newtonian approximation, 
depending only on the radial coordinate. From a physical point of view, such 
a function could be related to second order effects related to anisotropies 
and non-homogeneities  which could trigger the formation and the evolution 
of stellar systems. To take into account such a situation, one should perform 
the post-Newtonian limit of the theory
not only in the simple hypothesis of homogeneous spherical symmetry 
(Schwarzschild solution) but also considering more realistic situation as 
Lema$\hat{i}$tre--Tolman--Bondi solutions (see e.g.   
\citealt{herrera}).




\section{Discussion and Conclusions}

There is a growing attention to alternative model to the
$\Lambda$CDM paradigm as the latter is still suffering some
discrepancies at the galaxy scales and, most importantly, is
based on the assumption of the existence of two ingredients (DM
and DE) whose nature is still unknown.

Different attempts have tried to circumvent the problem by
introducing a modified dynamics, e.g. with the MOND theory (see
\citealt{2002ARA&A..40..263S}, \citealt{2010ApJ...718..380S}, \citealt{2010arXiv1011.5741C}), 
but this seems still needing some
DM at least to cluster scales which might be still consistent with
the primordial nucleosynthesis (e.g. via high energy neutrinos,
\citealt{2010MNRAS.402..395A}) and does not provide an explanation
for the DE.

Lately $f(R)$--gravity  models have made their step out as a
natural explanation for the two {\em dark} ingredients of the
Universe assuming that they are related to the fact that gravitational interaction could present further degrees of freedom whose dynamical effects emerge at large scales (\citealt{capozzfara}).
 In this paper we have checked the 
Yukawa-like modification to the Newtonian potential obtained
as post-Newtonian approximation of $f(R)-$gravity
for which no particular choice of the Lagrangian has been 
provided, with the only assumption that $f(R)$ is analytic function.

We have used a combination of long-slit spectroscopy and
planetary nebulae kinematics out to $\sim$7 \Re\ in three systems
(NGC 3379, NGC 4374, NGC 4494)
for which $\Lambda$CDM models
turned out to be fairly consistent with WMAP5 measurements (see
N+11 for a discussion).

Due to the small galaxy sample, the spirit of this analysis has
been to check whether $1)$ the modified potential introduced by the
$f(R)$-gravity allowed a fit to the galaxy kinematics
comparable to the DM models; $2)$ the three galaxies returned a
parameter $\delta$ which is comparable with spiral galaxies 
(\citealt{1984A&A...136L..21S}). 

We have found that the modified potentials allow to nicely model
the three galaxies with a distribution of the $\delta$ parameters
which turned out to be inconsistent with the
results found in spiral systems. We have shown some hints that $\delta$
might be correlated with the galaxy anisotropy, $\beta$, and the
scale parameter, $L$, with elliptical and spiral galaxies
following the same pattern.

This evidence can have interesting implication on the ability of the theory
to make predictions on the internal structure of the gravitating systems  
after their spherical collapse (e.g. \citealt{1977ApJ...218..592G}) which has 
to be confirmed on a larger galaxy sample 
which we expect to do in a near future.

Despite of some simplifications on the model adopted (e.g. constant
$M/L$ and anisotropy across the galaxy) and the degeneracies
between the model parameters, the results are very encouraging.
The fit to the data is very good in all cases and both the stellar
$M/L$ (with Kroupa IMF generally favored) and orbital anisotropy
turn out to be similar to the one estimated if a dark halo is
considered.

Getting a modified gravity to work self-consistently for all 
gravitating systems in general, and all galaxy families in particular, 
is a very non-trivial challenge that has foiled other theories (e.g. MOND).



\acknowledgments

We thank the anonymous referee for constructive comments which allowed us to 
significantly improve the paper. AJR was supported by National Science Foundation 
Grants AST-0808099 and AST-0909237. 
CT was supported by the Swiss National Science Foundation.


\end{document}